\documentclass[aps,twocolumn,longbibliography,pra,superscriptaddress,amssymb,amsmath,amsmath,floatfix]{revtex4-2}
\usepackage{graphicx}
\usepackage{epstopdf}
\usepackage{comment}
\usepackage{bm}
\usepackage{wrapfig}
\usepackage{times}
\usepackage[colorlinks=true,linkcolor=blue,urlcolor=blue,citecolor=blue]{hyperref}
\usepackage{float}
\begin{document}

\title{CaF+CaF interactions in the ground and excited electronic states: implications for collisional losses}

\author{Dibyendu Sardar}
\affiliation{Faculty of Physics, University of Warsaw, Pasteura 5, 02-093 Warsaw, Poland}
\affiliation{JILA, University of Colorado, Boulder, Colorado 80309, USA}

\author{Marcin Gronowski} 
\affiliation{Faculty of Physics, University of Warsaw, Pasteura 5, 02-093 Warsaw, Poland}

\author{Micha{\l} Tomza}
\email{michal.tomza@fuw.edu.pl}
\affiliation{Faculty of Physics, University of Warsaw, Pasteura 5, 02-093 Warsaw, Poland}

\author{John L. Bohn}
\email{bohn@murphy.colorado.edu}
\affiliation{JILA, University of Colorado, Boulder, Colorado 80309, USA}

\date{\today}

\begin{abstract}
Accurate \textit{ab initio} potential energy surfaces are essential to understand and predict collisional outcomes in ultracold molecular systems. In this study, we explore the intermolecular interactions between two laser-cooled CaF molecules, in both their ground and excited electronic states, aiming to understand the mechanisms behind the observed collisional losses on the nonreactive, spin-polarized surface of the CaF+CaF system. Using state-of-the-art \textit{ab initio} methods, we compute 12 electronic states of the Ca$_2$F$_2$ complex within the rigid rotor approximation applied to CaF. Calculating the potential energy surfaces for the excited electronic states of Ca$_2$F$_2$ is challenging and computationally expensive. Our approach employs the multireference configuration interaction method, restricted to single and double excitations, along with a reasonably large active space to ensure the convergence in the excited states. We also compute the spin-orbit coupling between the ground state and the lowest spin-polarized triplet state, as well as the spin-spin coupling within the lowest triplet state (1)~$^3\mathrm{A}'$. Additionally, we determine the electric transition dipole moments for the (1)~$^3\mathrm{A}'$--(2)~$^3\mathrm{A}'$ and (1)~$^3\mathrm{A}'$--(1)~$^3\mathrm{A}''$ transitions. Notably, we find that the lowest spin-polarized state (1)~$^3\mathrm{A}'$, shifted by 1064 nm of laser light from the optical dipole trap, intersects several electronically excited states. Finally, by analyzing the potential energy surfaces, we discuss two plausible pathways that may account for the observed collisional losses on the spin-polarized surface of the CaF+CaF system. 
\end{abstract}

\maketitle

\section{Introduction}
\label{sec:intro}
Ultracold polar molecules constitute an excellent system for studying fundamental quantum physics and chemistry \cite{CarrNJP2009}. The rich internal structure and controllable intermolecular interactions make them an ideal candidate for a wide range of applications, from quantum computations \cite{DemillePRA2002,YelinPRA2006} and quantum simulations \cite{CornishNatPhys2024} of many-body systems to precession measurements \cite{DemilleScience2017} for fundamental physics. Additionally, ultracold molecules offer the possibility to study chemical reactions with an unprecedented level of control \cite{BellMOLPHYS2009,KarmanNP24,LiuNAT2021}. These systems are particularly fascinating due to the presence of inherent long-range anisotropic dipole-dipole interactions, which can be controlled using externally applied fields \cite{GadwayJPB2016}.

Ultracold heteronuclear molecules in their absolute ground state are obtained in two ways: the association method and the direct laser cooling technique. The association scheme depends on the two steps, initially a pair of ultracold atoms is magnetoassociated by ramping the magnetic field to form a weakly bound Feshbach molecule \cite{ChinRMP2010}, followed by coherent optical transfer to the absolute ground state by stimulated Raman adiabatic passage \cite{KralRMP2007,BergmannMRP1998}. The technique has paved the way for the creation of alkali-metal dimers in their absolute singlet ground state (X~$^1\Sigma^+$) \cite{VogesPRL2020,YangPRL2020,GuoPRL2016,ParkPRL2015,NiScience2008,MolonyPRL2014,StevensonPRL2023,ZamarskiARXIV2025} as well as in the lowest triplet electronic state (a~$^3\Sigma^+$) \cite{RvachovPRL2017}. Alternatively, direct laser cooling relies on highly diagonal Franck–Condon factors and has been demonstrated for certain classes of molecules, including alkaline-earth monofluorides and oxides such as CaF \cite{AndereggNATPHYS2018,CheukPRL2018,CaldwellPRL2019}, SrF \cite{ShumanNAT2010}, and YO \cite{DingPRX2020}.

Collisions between ultracold polar molecules often result in loss, typically exhibiting the characteristics of a two-body process. In most cases, the observed loss rates approach the universal limit, meaning that almost every collision leads to the loss of both molecules \cite{IdziaszekPRL2010}. For chemically reactive species, this loss could be attributed to exothermic chemical reactions \cite{OspelkausScience2010}. However, even in the case of chemically nonreactive molecules, substantial loss has been observed \cite{ParkPRL2015}. This is believed to arise from the formation of long-lived collision complexes, which can absorb photons from the trapping light (e.g., an optical dipole trap), leading to photoexcitation and subsequent loss from the trap \cite{LiuNat.Phys2020,GregoryPRL2020}. Although the theoretical model \cite{ChristianenPRL2019} successfully explains the observed loss in certain alkali-metal dimers \cite{BausePRL2023}, it does not account for the loss behavior in other classes of alkali molecules \cite{BausePRX2021,GersemaPRL2021}. Alongside, collisions of ultracold paramagnetic alkaline-earth fluoride molecules, such as CaF+CaF \cite{SardarPRA2023} or SrF+SrF \cite{MeyerPRA2011}, show chemical reactivity in the electronic ground state. Again, for collisions in their lowest spin-polarized state, the observed loss approaches the universal limit \cite{CheukPRL2020}. 

The CaF presents an interesting case where multiple loss mechanisms may occur. Spin-polarized CaF molecules will encounter each other on a triplet surface, on which chemical reactivity is endothermic and cannot occur \cite{SardarJPCA2023}. However, even on this surface, three loss possibilities compete: 1) the molecules could simply form a complex in the triplet potential and vanish from observation. 2) The molecules in the triplet complex could absorb photons from the optical dipole trapping light and be subsequently destroyed at a rate dependent on the intensity of this light. 3) A nonadiabatic transition could occur, due to a direct triplet-singlet crossing and spin-orbit coupling, which transfers the molecules to the singlet ground state, which is in fact chemically reactive \cite{SardarPRA2023}.

To understand the interplay of these possible loss mechanisms requires a detailed understanding of the intermolecular interactions between two molecules in both their ground and excited electronic states. In this context, comprehensive calculations of the ground and excited state potential energy surfaces, and transition dipole moments, have so far been carried out only for the NaK+NaK system \cite{ChristianenJCP2019,ChristianenPRL2019}, and have been used to investigate the observed two-body collisional losses. Additionally, full-dimensional potential energy surfaces have been constructed for the KRb+KRb \cite{YangJPCL2020,HuangJPCA2021} and NaRb+NaRb \cite{LiuJCPA2022} systems, limited to the electronic ground state. Furthermore, electronic ground-state interactions have been studied and analyzed for pairs of alkaline-earth monofluoride molecules, such as CaF+CaF \cite{SardarPRA2023,SardarJPCA2023} and SrF+SrF \cite{MeyerPRA2011}. However, to the best of our knowledge, excited electronic states, either of CaF+CaF or other similar systems, remain unexplored both theoretically and experimentally.    

Here, we fill this gap and investigate the intermolecular interactions between two CaF molecules in their excited electronic states. The primary objective of this study is to accurately calculate the molecular potential energy surfaces for the CaF+CaF system and to discuss plausible pathways for the observed losses on the nonreactive and spin-polarized surface of CaF+CaF. Our study focuses on the ultracold CaF molecule, which is amenable to magnetic and optical trapping, and can be cooled to temperatures as low as a few microkelvins \cite{AndereggNATPHYS2018,CheukPRL2018,CaldwellPRL2019}. In addition, recent experiments have explored ultracold collisions between CaF molecules confined in optical tweezers \cite{CheukPRL2020}.

The interaction between two CaF molecules in their electronic ground states results in the formation of two molecular states: a ground singlet (1)~$^1\mathrm{A}'$ and a spin-polarized triplet (1)~$^3\mathrm{A}'$, analogous to the interaction between two alkali-metal atoms \cite{LadjimiPRA2024}. However, when one of the CaF molecules is electronically excited, the interaction becomes significantly more complex, giving rise to multiple potential energy surfaces of various symmetries and spin multiplicities. While the ground-state potential energy surfaces for the CaF+CaF system have been previously reported \cite{SardarPRA2023,SardarJPCA2023}, obtaining accurate excited-state surfaces remains a considerable challenge. This difficulty arises primarily from the complexity of the four-body system and the inherent limitations in converging excited-state solutions using \textit{ab initio} methods. In the present work, we address this issue by carefully selecting a sufficiently large active space in combination with a quadruple-zeta basis set, aiming to achieve a reliable and accurate description of the excited states of the Ca$_2$F$_2$ dimer. 

In this study, we compute one-dimensional cuts of the potential energy surfaces (PESs) for the 12 electronic states arising from the three lowest dissociation channels of the CaF+CaF system. These dissociation channels correspond to an excitation energy of up to 19000~cm$^{-1}$ in the CaF monomer. The electronic structure calculations are performed using the multireference configuration interaction method, restricted to single and double excitations, under the rigid-rotor approximation for the CaF monomer. We find that the influence of molecular nonrigidity on the shape of the PES for the lowest spin-polarized state, (1)~$^3\mathrm{A'}$, is negligible. Notably, the PES of the (1)~$^3\mathrm{A'}$ state, when shifted by 1064 nm laser light from an optical dipole trap, intersects multiple excited-state potential energy surfaces. The excited states of the Ca$_2$F$_2$ dimer are strongly bound and exhibit substantial interactions, and many of them show avoided crossing or conical intersections between the same symmetry states. Therefore, it is entirely plausible that photoexcitation during collisions can account for the loss in optical dipole traps.


We also compute the spin-orbit coupling between the ground singlet (1)~$^1\mathrm{A}'$ state and the lowest triplet (1)~$^3\mathrm{A}'$ state, and the spin-spin coupling within the (1)~$^3\mathrm{A}'$ state. In addition, we calculate the electric transition dipole moments for selected dipole-allowed transitions, $(1)~^3\mathrm{A}'–(2)~^3\mathrm{A}'$ and $(1)~^3\mathrm{A}'–(1)~^3\mathrm{A}''$, respectively, and these transitions are expected to be important in calculating the rate of photoexcitation. By analyzing the potential energy surfaces along with the computed transition dipole moments and spin-dependent couplings, we discuss plausible pathways that may contribute to the observed two-body losses for CaF+CaF collisions on the spin-polarized surface.

The structure of the paper is the following. In Sec.~\ref{sec:methods} we describe the employed computational methods for the \textit{ab initio} calculations. In Sec~\ref{sec:results} we present and discuss the properties of the molecular potential energy surfaces, transition dipole moments, and spin-orbit and spin-spin couplings. In Sec~\ref{sec:summary} we provide a summary and outlook.

\section{METHODS OF CALCULATIONS}
\label{sec:methods} 
\begin{table*}
\caption{\label{tab:channel}Possible molecular states of the lowest four asymptotic channels of CaF+CaF under the C$_\mathrm{s}$ symmetry, where $\Delta E$ is the energy difference between the dissociation thresholds with regard to the ground state asymptote~\cite{KramidaNIST2018}. The value in parentheses indicates the number of molecular states.}
\begin{ruledtabular}
\begin{tabular}{ccc}
 Dissociation threshold  & Molecular states of Ca$_2$F$_2$&$\Delta E$ (cm$^{-1}$)\\
\hline
 CaF$(\mathrm{X}~^2\Sigma^+)$+CaF$(\mathrm{X}~^2\Sigma^+)$ & $^1$A$'$, $^3$A$'$  & 0 \\ 
 CaF$(\mathrm{X}~^2\Sigma^+)$+CaF$(\mathrm{A}~^2\Pi)$ & $^1$A$'$ ($\times 2$), $^3$A$'$ ($\times 2$), $^1$A$''$ ($\times 2$), $^3$A$''$ ($\times 2$)  & 16490 \\ 
 CaF$(\mathrm{X}~^2\Sigma^+)$+CaF$(\mathrm{B}~^2\Sigma^+)$ & $^1$A$'$ ($\times 2$) , $^3$A$'$ ($\times 2$)  & 18844 \\ 
 CaF$(\mathrm{X}~^2\Sigma^+)$+CaF$(\mathrm{C}~^2\Pi)$ & $^1$A$'$ ($\times 2$), $^3$A$'$ ($\times 2$), $^1$A$''$ ($\times 2$), $^3$A$''$ ($\times 2$) & 30255 \\ 
\end{tabular}
\end{ruledtabular}
\end{table*}

We employ \textit{ab initio} quantum chemical methods to understand the interactions between two CaF molecules both in their ground and excited electronic states. The molecular potential energy surfaces are calculated using the \textsc{Molpro} 2022.1 software package \cite{WernerJCP2020}.  

Calcium monofluoride is a paramagnetic molecule with an unpaired electron. The electronic ground state of CaF is denoted by X~$^2\Sigma^+$, while the first excited state is represented by A~$^2\Pi$. When two CaF molecules interact in their ground state, they form the (1)~$^1\mathrm{A}'$ and (1)~$^3\mathrm{A}'$ states, similar to alkali-metal dimers. However, when one molecule is in the ground state and the other in the first excited state, the interactions give rise to the $\mathrm{A}'$ and $\mathrm{A}''$ states, with singlet and triplet spin multiplicities. Table~\ref{tab:channel} presents the molecular states of CaF+CaF for the four lowest dissociation channels, together with the energy differences relative to the ground-state asymptote.

For the calculation of the molecular PESs of CaF+CaF, we use the augmented correlation-consistent polarized weighted core-valence quadrupole-$\zeta$ quality basis set, specifically aug-cc-pwCVQZ-PP for calcium \cite{HillJCP2017}  and aug-cc-pwCVQZ for fluorine \cite{HillJCP2017}. To account for scalar relativistic effects in Ca, we employ the small-core relativistic energy-consistent pseudopotential (ECP10MDF) \cite{LimJCP2006} to replace the 10 inner-shell electrons of Ca. We determine the interaction energy in CaF+CaF using the supermolecular method, applying the Boys-Bernardi counterpoise correction to account for the basis set superposition error.

Accurate calculations of the excited electronic states of the Ca$_2$F$_2$ dimer are particularly challenging and require the careful selection of an appropriate active space within the framework of multireference electronic structure methods. To this end, we construct the active space for the excited states of Ca$_2$F$_2$ using the following systematic procedure. Calcium monofluoride, under the consideration of the effective core potential (ECP) of Ca, has 19 electrons occupying the 10 lowest-energy molecular orbitals. Therefore, the minimal active space for the CaF dimer consists of 38 electrons occupying 19 molecular orbitals for the ground (1)~$^1\mathrm{A}'$ state, and 20 orbitals for the lowest spin-polarized triplet (1)~$^3\mathrm{A}'$ state. To describe the excited states of Ca$_2$F$_2$, we incorporate the 3d, 4p, and 5s atomic orbitals of each Ca atom, resulting in an expansion of the active space to 38 molecular orbitals. Within this active space, correlation energies for all 38 electrons can be computed; however, such a calculation is highly demanding in terms of computational resources. Therefore, we consider an optimal active space by selectively reducing the electron correlation while maintaining the convergence in the excited states. Specifically, we consider the lowest 12 molecular orbitals, derived from the deeply bound orbitals of Ca$_2$F$_2$, as frozen core orbitals. Additionally, we treat the next 6 orbitals as closed-shell and allocate the remaining 20 as open-shell orbitals. Consequently, we account for the correlation among 14 electrons within 26 orbitals, ensuring computational efficiency while providing a reliable description of the excited electronic states of Ca$_2$F$_2$.

We use an internally contracted multireference configuration interaction method restricted to single and double excitations, MRCISD \cite{ShamasundarJCP2011,WernerJCP1988} to calculate the one-dimensional cuts and two-dimensional PESs for both ground and excited electronic states of CaF+CaF, including spin-orbit interaction.  First, we perform a complete active space self-consistent field (CASSCF) calculation on the mentioned active space. The resulting CASSCF wavefunction serves as a reference for the subsequent MRCISD calculation. The MRCISD calculation includes a Davidson correction, which approximately accounts for the size consistency. Additionally, we employ the coupled cluster method restricted to single, double, and perturbative triple excitations CCSD(T) \cite{KnowlesJCP2000,KnowlesJCP1993} to calculate the lowest electronic states of the Ca$_2$F$_2$ dimer within the C$_\mathrm{s}$ and C$_{\mathrm{2v}}$ symmetries.

Additionally, the spin-orbit coupling between the two lowest electronic states of Ca$_2$F$_2$,  (1)~$^1\mathrm{A}'$-(1)~$^3\mathrm{A}'$, is evaluated by using the \textsc{Molpro} package. First, a state-averaged complete active space self-consistent field (SA-CASSCF) calculation is performed to obtain reference wavefunctions. Next, dynamical correlation effects are incorporated using the MRCISD calculations. The spin-orbit coupling (SOC) matrix elements are then evaluated using the ECP-LS approach \cite{BerningMolPhys2000}, which includes relativistic spin-orbit interactions directly within the effective core potential. Finally, the SOC matrix is diagonalized to obtain eigenstates and energy splittings, which provide a detailed understanding of spin-orbit effects in the Ca$_2$F$_2$ dimer.


We also calculate the electronic spin-spin coupling using the bias-corrected spin-dependent second-order Dynamic Correlation Dressed Complete Active Space [DCD-CAS(2)] method~\cite{RemiJCTC2009,PathakJCP2017,KollmarJCC2019,LangJCP2019,UgandiIJQC2023}. All electrons are included in the calculation, while the active space is restricted to two electrons in eight orbitals. Relativistic effects are treated using the one-electron version of the exact two-component (X2C) Hamiltonian \cite{PengJCP2013}. We tested several all-electron relativistic Karlsruhe-type basis sets and found that the basis set size has only a moderate influence on the predicted spin-spin coupling. Therefore, we report results obtained with the largest basis set employed: x2c-QZVPPall-s \cite{FranzkePCCP2019,PollakJCTC2017}. All calculations in this part are performed using the \textsc{Orca} 6.0 program \cite{NeeseJCP2020,NeeseJCC2023}.

\begin{figure}[b]
\includegraphics[width=0.30\textwidth]{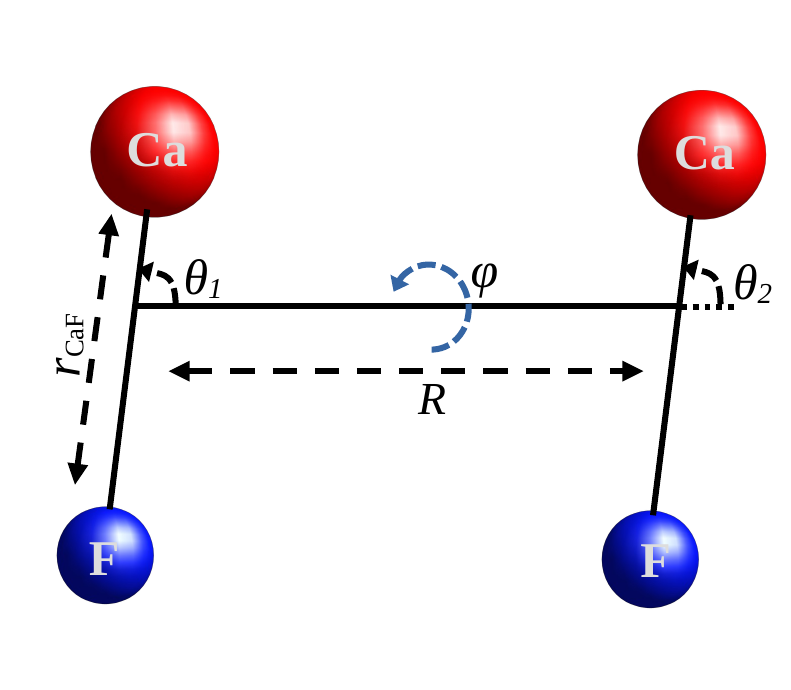}
\caption{\label{fig:schematic} Schematic diagram for the molecular orientations of CaF+CaF in the Jacobi coordinates.}
\end{figure}

The electronic structure calculations for potential energy surfaces are performed for various configurations, including one-dimensional (1D) and two-dimensional (2D) cuts, within the rigid-rotor approximation in which the CaF bond length is held fixed. In representing the geometry of the CaF+CaF tetratomic system, we use the Jacobi coordinates to define the relative arrangement of the CaF molecules as shown in Fig.~\ref{fig:schematic}. In this framework, $R$ represents the distance between the centers of mass of the two CaF monomers, each of which has the bond length $r_{\mathrm{CaF}}$. The molecular orientations are characterized by the polar angles $\theta_1$ and $\theta_2$ relative to the intermolecular axis, while the torsional angle is described by $\varphi$. In our \textit{ab initio} electronic structure calculations, we treat the CaF as a rigid rotor, fixing its bond length at the experimentally determined equilibrium value of $r_{\mathrm{CaF}}=3.695$~bohr \cite{KaledinJMSP1999}. However, we also investigate the effects of bond-length flexibility in our calculations to assess the impact of nonrigidity on the interaction energy of the system.

\begin{figure*}
\includegraphics[width=\linewidth]{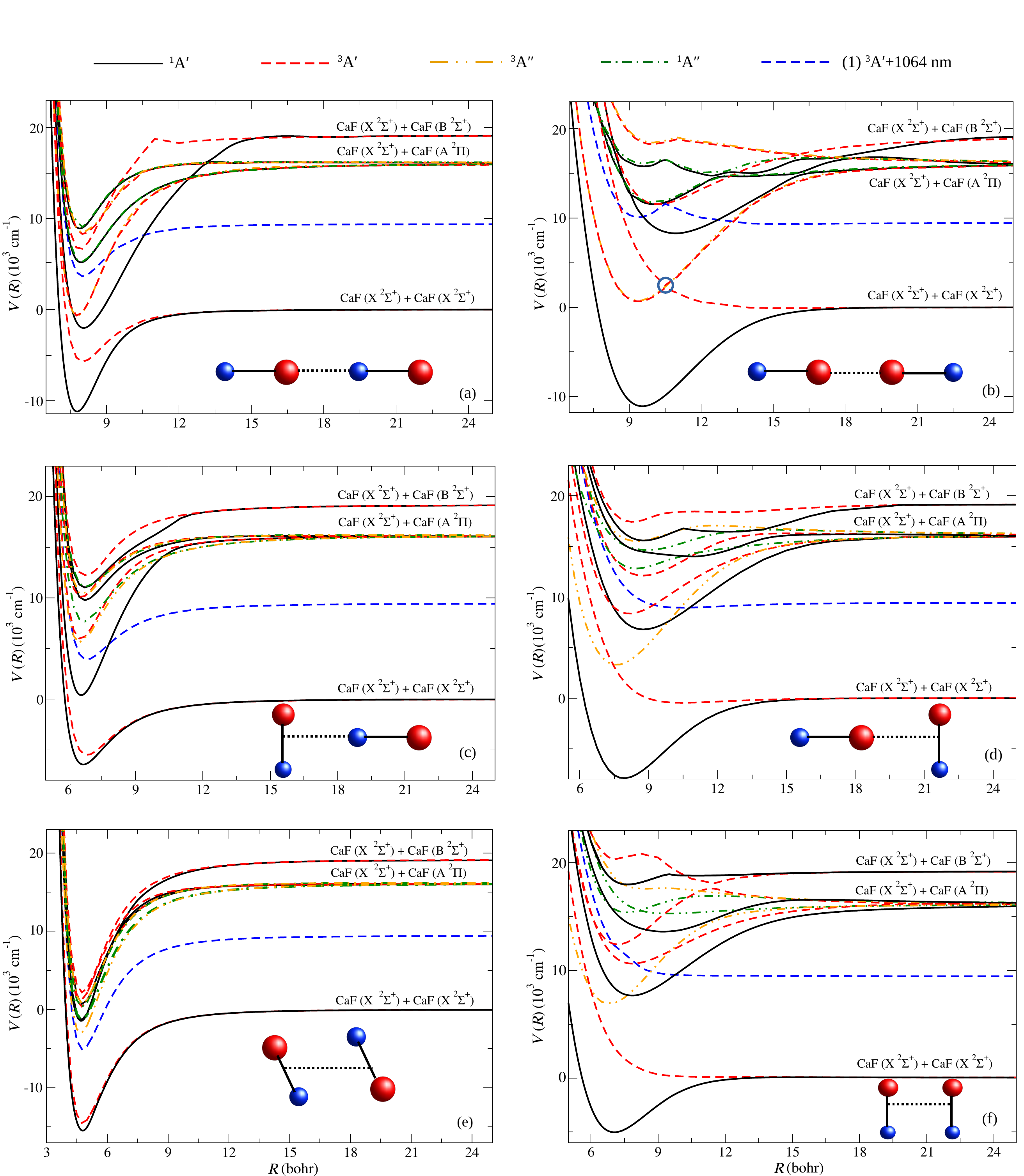}
\caption{\label{fig:all} One-dimensional cuts of potential energy surfaces as a function of $R$ for six orientations of CaF+CaF: two linear configurations in panels (a) and (b), two T-shaped configurations in panels (c) and (d), and two parallel configurations in panels (e) and (f), where the former includes a global minimum geometry. The solid black and red dashed curves represent the $^1\mathrm{A}'$ and $^3\mathrm{A}'$ states, respectively. The $^1\mathrm{A}''$ and $^3\mathrm{A}''$ states are shown as green dash-dotted and orange dash-double-dotted curves, respectively. The blue dashed curve represents the PES of the (1)~$^3\mathrm{A}'$ state shifted by the energy of the photon of 1064~nm wavelength. The circle in panel (b) presents a conical intersection, which is further discussed and analyzed in the context of the validity of the rigid rotor approximation in PES calculations for CaF+CaF.}
\end{figure*}

\section{RESULTS AND DISCUSSION}
\label{sec:results}
\subsection{Equilibrium properties of CaF and Ca$_2$F$_2$}
\label{subsec:eqgeom}
We investigate the spectroscopic parameters for the two lowest electronic states of CaF using the specified basis sets and methods, and compare our findings with the existing literature. In Table~\ref{tab:monomer}, we present the equilibrium bond length ($r_\mathrm{CaF}$) and the depth of the well ($D_e$) for the X~$^2\Sigma^+$ and A~$^2\Pi$ states of the CaF molecule, calculated by the CCSD(T) and MRCISD methods. The calculated spectroscopic parameters from both methods agree well with literature-reported values with an error of less than 1\%. Additionally, the permanent electric dipole moment, obtained from the MRCISD calculation, agrees well with the available experimental value for the ground X~$^2\Sigma^+$ state, as well as with the CCSD(T) result for the excited A~$^2\Pi$ state of CaF.

Next, we calculate the potential well depths at the global minima for the two lowest (1)~$^1\mathrm{A}'$ and (1)~$^3\mathrm{A}'$ molecular states of the Ca$_2$F$_2$ dimer. The geometrical parameters for the global minimum of these two states are discussed in detail in the reference \cite{SardarJPCA2023}. The global minimum of both of these states exhibits $\mathrm{D}_{\mathrm{2h}}$ symmetry with similar geometrical parameters and comparable well depth, unlike alkali-metal dimers \cite{LadjimiPRA2024}. This unusual feature of the Ca$_2$F$_2$ dimer originates from the ionic nature of its bonding. In the global geometry, two calcium atoms separated by 6.4~bohr each carry a partial positive charge and possess an unpaired electron. Due to the near-zero orbital amplitude at the midpoint, the distinction between bonding and antibonding orbitals is minimized, resulting in a similar optimized minimum for both states, (1)~$^1\mathrm{A}'$ and (1)~$^3\mathrm{A}'$. In our calculations, we use correlation-consistent quadruple-$\zeta$ basis sets and the multireference configuration interaction method restricted to single and double excitations. The resulting potential well depths for the ground (1)~$^1\mathrm{A}'$ state and the lowest spin-polarized (1)~$^3\mathrm{A}'$ state are 17979~cm$^{-1}$ and 17476~cm$^{-1}$, respectively. These results are in good agreement and consistent with previous findings reported in Refs. \cite{SardarPRA2023,SardarJPCA2023}.

\begin{table}[tb!]
\caption{\label{tab:monomer}
Equilibrium properties for the lowest two electronic states of CaF, including equilibrium bond length (bohr), depth of the well (cm$^{-1}$), and norm of the permanent electric dipole moment $|d_{\mathrm{CaF}}|$ ($e a_0$). The dipole moment values are computed through numerical differentiation of the electronic energy with respect to an applied electric field.}
\begin{ruledtabular}
\begin{tabular}{ c c c c c c } 
State & Method & $r_\text{CaF}$ & $D_e$ & $|d_{\mathrm{CaF}}|$  \\
\hline
X~$^{2}\Sigma^+$ 
& CCSD(T) & 3.69 & 44254 & 1.218  \\ 
& MRCI+Q& 3.69 & 44485 & 1.196 \\
& MRCI  & 3.69 & 43408      & 1.142 \\
& MRCI+Q \cite{FernandesSPA2020}&3.718 & 44602 & \\
& RCCSD(T) \cite{HouJQSRT2018}                &      &       & 1.196 \\
& Exp. \cite{CharronJMS1995,HouJQSRT2018,ChildsJCP1984,ChildsJMS1986} & 3.6880 & 44203.5 & 1.208  \\ 

A~$^{2}\Pi$ 
& CCSD(T) & 3.66 & 27481  & 1.014  \\
& MRCI+Q  & 3.66 & 27868  & 1.005 \\
& MRCI    & 3.66     & 27428       & 0.997 \\
& MRCI+Q \cite{FernandesSPA2020}&3.656 & 27584 \\
& Exp. \cite{KaledinJMSP1999} & 3.6612 &  & \\ 
\end{tabular}
\end{ruledtabular}
\end{table}

\subsection{1D PESs of CaF+CaF}
\label{subsec:1dcut}
We calculate the one-dimensional cuts of the potential energy surface for the ground and excited electronic states, considering the excitation energy up to approximately 19000~cm$^{-1}$ of the CaF monomer. Three dissociation channels of the dimer Ca$_2$F$_2$ lie within this excitation energy range, while a fourth dissociation channel lies at a much higher energy and is likely irrelevant for experimental studies. It is important to mention that the second dissociation channel, CaF(X~$^2\Sigma^+$)$+$CaF$(\mathrm{A}~^2\Pi)$, yields two $\mathrm{A}'$ and two $\mathrm{A}''$ states with both singlet and triplet characteristics. This arises from the symmetry of the molecular approach: when one CaF (X~$^2\Sigma^+$) molecule approaches from the left and another CaF (A~$^2\Pi$) from the right, their interaction yields one $\mathrm{A}'$ and one $\mathrm{A}''$ state. A reversed approach (right-left) results in an identical number of states due to the exchange interaction. As a result, there are four spatial symmetry states—two $\mathrm{A}'$ and two $\mathrm{A}''$. Each of these spatial symmetry states can exist in both singlet and triplet spin configurations, resulting in a total of eight electronic states associated with the second dissociation channel of the Ca$_2$F$_2$ dimer (Table~\ref{tab:channel}). Similarly, the third dissociation channel of the Ca$_2$F$_2$ dimer yields a total of four molecular states. In our calculations, we include only two of these states, one $^1\mathrm{A}'$, and one $^3\mathrm{A}'$,  due to the convergence issues.

These excited states of the dimer are useful to provide valuable insight for the observed collisional loss mechanisms on the spin-polarized surface of CaF+CaF. The collision between two CaF molecules on the spin-polarized surface may lead to the formation of a collision complex, which can undergo excitation to the nearby electronic excited states of the Ca$_2$F$_2$ dimer by absorbing photons from the trapping laser, resulting in a trap loss. Additionally, another mechanism for collisional loss on the spin-polarized surface of CaF+CaF, may be attributed by spin-orbit-mediated transitions from the nonreactive surface, (1)~$^3\mathrm{A}'$, to the reactive ground state, (1)~$^1\mathrm{A}'$, where the excited states of the Ca$_2$F$_2$ dimer could play a crucial role. Particularly, the admixture of the electronically excited states to the ground state may enhance the probability of such a transition, thereby increasing the trap loss of the CaF molecule.


We compute one-dimensional cuts of the potential energy surfaces to provide a simplified yet insightful view of the interactions of the Ca$_2$F$_2$ dimer, without computing complex full-dimensional surfaces for the excited states. Still, we encounter some convergence problems, especially at repulsive walls, that may lead to the small discontinuities in PESs of high-lying excited states. However, these are irrelevant for studying collisional loss. We calculate 1D-PES for the 12 lowest electronic states under the rigid rotor approximation, using the MRCISD method. In particular, we consider four states for both the $^1\mathrm{A}'$ and $^3\mathrm{A}'$ symmetries, as well as two states each for the $^1\mathrm{A}''$ and $^3\mathrm{A}''$ symmetries, corresponding to different orientations of the CaF+CaF system. Our calculations account for two linear orientations ($\theta_1=0^\circ, \theta_2=0^\circ; \text {and} \hspace{0.1in}  \theta_1=0^\circ, \theta_2=180^\circ$), two T-shaped geometries ($\theta_1=90^\circ, \theta_2=0^\circ; \text {and} \hspace{0.1in}  \theta_1=0^\circ, \theta_2=90^\circ$), and two parallel orientations ($\theta_1=90^\circ, \theta_2=90^\circ$ and $\theta_1=121^\circ, \theta_2=59^\circ, \varphi=180^\circ$, where the latter includes a global minimum geometry).

In Fig.~\ref{fig:all}, we present one-dimensional cuts of the potential energy surfaces for the six different orientations of CaF+CaF. The position of the minimum for the ground state, (1)~$^1\mathrm{A}'$, varies with orientations, however, the order of magnitude for the depth of the well remains the same. In contrast, the (2)~$^1\mathrm{A}'$ state exhibits the deepest well for the linear configuration, as shown in Fig. \ref{fig:all}(a). For this state, a small portion of the conformational space lies below the dissociation limit of CaF$(\mathrm{X}~^2\Sigma^+)$+CaF$(\mathrm{X}~^2\Sigma^+)$ for both the linear and global minimum geometries, illustrated in Figs.~\ref{fig:all}(a) and \ref{fig:all}(e), respectively. On the other hand, for the lowest triplet state, (1)~$^3\mathrm{A}'$, the PES shows attractive interactions for linear, T-shaped, and global minimum geometries [Figs.~\ref{fig:all}(a), \ref{fig:all}(c), and \ref{fig:all}(e)], but is repulsive for the remaining orientations. Notably, the PES of the first excited triplet state, (2)~$^3\mathrm{A}'$, approaches that of the (1)~$^3\mathrm{A}'$ state only in the linear configuration as shown in Fig.~\ref{fig:all}(b). These findings highlight how variations in the electronic state of CaF significantly influence the shape of the potential energy surface in the Ca$_2$F$_2$ dimer.

By analyzing the potential energy surfaces, we find that the excited states of Ca$_2$F$_2$ are strongly bound, with potential wells that are, in some cases, deeper or comparable to that of the ground state, (1)~$^1\mathrm{A}'$. The density of electronic states increases significantly at higher excitation energies. Additionally, the excited states exhibit strong interactions, with many showing avoided crossings between states of the same symmetry, suggesting the presence of strong non-adiabatic radial couplings between the states.

In the CaF+CaF system, we observe several avoided crossings and conical intersections (Fig.~\ref{fig:all}) between electronic states of identical symmetry. For linear orientation ($\theta_1 = 0^\circ, \theta_2 = 0^\circ$), the (2)~$^1\mathrm{A}'$ state exhibits an crossing with the (3)~$^1\mathrm{A}'$ state around $R \approx 12.5$~bohr, followed by another crossing between the (3)~$^1\mathrm{A}'$ and (4)~$^1\mathrm{A}'$ states near $R \approx 13$~bohr. Additional crossings are observed in the pair (2)~$^1\mathrm{A}'$–(3)~$^1\mathrm{A}'$ at approximately $R \approx 8.8$ and $16.2$~bohr for alternative linear geometries, along with an crossing between the states (3)~$^1\mathrm{A}'$ and (4)~$^1\mathrm{A}'$ around $R \approx 12.9$~bohr. However, these crossings disappear when the dimer becomes bent. In T-shaped configurations, avoided crossings are detected between the (3)~$^1\mathrm{A}'$ and (4)~$^1\mathrm{A}'$ states at $R \approx 8.2$~bohr and $R \approx 14.5$~bohr, as shown in Figs~\ref{fig:all}(c) and \ref{fig:all}(d), respectively. Similarly, for states of the $^3\mathrm{A}'$ symmetry, the excited (2)~$^3\mathrm{A}'$ state approaches the lowest spin-polarized (1)~$^3\mathrm{A}'$ state in the highly symmetric linear configuration of CaF+CaF, leading to a conical intersection, as depicted in Fig.~\ref{fig:all}(b). Upon a slight deviation from linearity, the molecular symmetry is reduced, resulting in an avoided crossing between these states. This conical intersection between the (1)~$^3\mathrm{A}'$ and (2)~$^3\mathrm{A}'$ states is analyzed in greater detail in the following subsection.

We note that the ground state, (1)~$^1\mathrm{A}'$, lies significantly lower in energy than the lowest triplet state, (1)~$^3\mathrm{A}'$, near the minima of the potential energy surfaces for most orientations. However, the two states come close in energy only in the T-shaped and global-minimum configurations of CaF+CaF, as shown in Figs~\ref{fig:all}(c) and \ref{fig:all}(e), respectively. Importantly, no direct crossing is observed between the states (1)~$^1\mathrm{A}'$-(1)~$^3\mathrm{A}'$ in any of the configurations studied in Fig.~\ref{fig:all} and several additional scans we did for these two states.

We find that the PESs for the (2)~$^1\mathrm{A}'$ (or (3)~$^1\mathrm{A}'$) state is energetically degenerate with (1)~$^1\mathrm{A}''$ (or (2)~$^1\mathrm{A}''$) for the linear configurations of CaF+CaF. This accidental degeneracy arises due to the presence of the degenerate $\Pi$ components in the C$_\mathrm{s}$ point group. However, the degeneracy is lifted for other nonlinear configurations of CaF+CaF. A similar conclusion can be applied for the pair states (2)~$^3\mathrm{A}'$-(1)~$^3\mathrm{A}''$ and (3)~$^3\mathrm{A}'$-(2)~$^3\mathrm{A}''$, respectively.

\begin{figure}
\includegraphics[width=\linewidth]{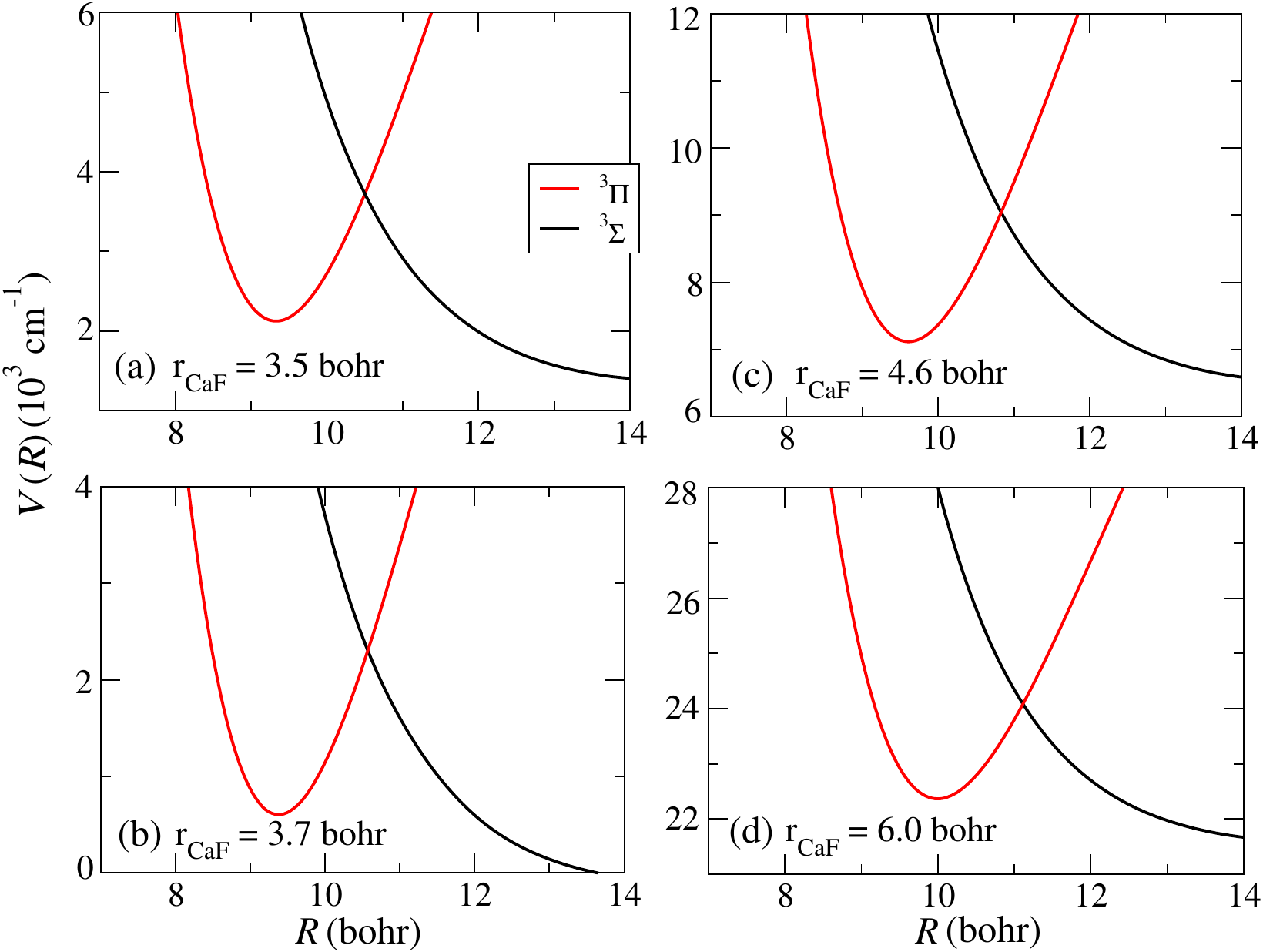}
\caption{Variation of interaction energy under rigid and nonrigid rotor approximations of CaF, evaluated for a specific CaF+CaF configuration ($\theta_1 = 0^\circ$, $\theta_2 = 180^\circ$). The solid black and red curves represent the lowest $^3\Sigma$ and $^3\Pi$ states, respectively, under the C$_{\mathrm{2v}}$ point group. Panel (b) shows the potential energy surfaces at the equilibrium bond length of CaF, while panels (a) and (c) depict PESs for bond compression and elongation, respectively. Panel (d) shows an additional elongation case. The interaction energy is calculated with respect to the ground state asymptote, CaF$(\mathrm{X}~ ^2\Sigma^+)$+CaF$(\mathrm{X}~ ^2\Sigma^+)$. Note different vertical scales in different panels. \label{fig:rigidity}}
\end{figure}

\subsection{Validity of rigid-rotor approximation of CaF molecule}
\label{subsec:rigidrotor}
The one-dimensional cut plots described above are computed under the rigid rotor approximation for the CaF molecule. To validate this approximation in the PES calculations, we further analyze the crossing between the pair of the (1)~$^3\mathrm{A}'$ and (2) $^3\mathrm{A}'$ states shown in Fig.~\ref{fig:all}(b). The interaction energies of these two states around the crossing are computed using \textit{ab initio} coupled cluster calculations. We consider three different values of monomer bond length, $r_{\text {CaF}}$, including the equilibrium geometry. The resulting PESs are presented in Fig.~\ref{fig:rigidity}. 

Under the C$_{2\mathrm{v}}$ symmetry, the lowest spin-polarized (1)~$^3\mathrm{A}'$ state correlates to a~$^3\Sigma$ state, while the (2)~$^3\mathrm{A}'$ state correlates to a $^3\Pi$ state, with a direct crossing at $R = 10.5$~bohr. We note that, whether the CaF bond length is elongated or compressed relative to its equilibrium value, the shape and topology of the PES for the $^3\Sigma^+$ state remains essentially unchanged. Moreover, this PES remains above the ground-state dissociation threshold. Additionally, the interaction energy near the minimum of the $^3\Pi$ state does not cross the ground-state asymptote, CaF$(\mathrm{X}~^2\Sigma^+)$+CaF$(\mathrm{X}~^2\Sigma^+)$. The crossing between the $^3\Sigma$ and $^3\Pi$ states is shifted to higher energy relative to the ground-state asymptote by a change of binding energy in CaF monomers. These observations suggest that deviations from the rigid rotor approximation have a negligible effect on the shape and the topology of available PESs of the Ca$_2$F$_2$ dimer, likely due to the substantial binding energy of the CaF monomer. Moreover, unlike in the KRb+Rb collisions \cite{LiuNatChem2025}, no significant enhancement of the spin-rotation interaction is expected in this case, as the conical intersection is energetically inaccessible.

\begin{figure}[t]
\includegraphics[width=\linewidth]{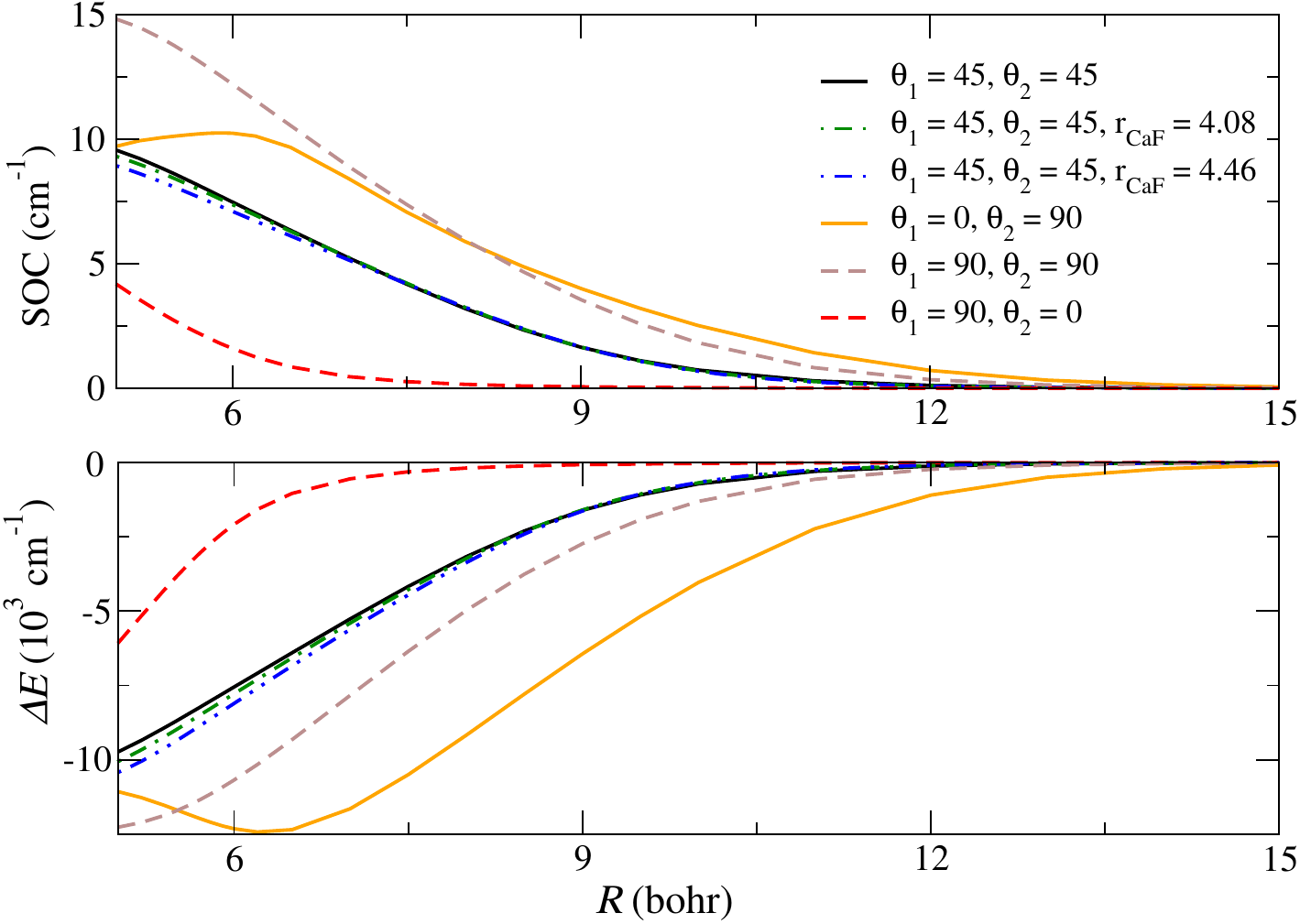}
\caption{Upper panel: variation of the spin-orbit matrix element between the two lowest electronic states, (1)~$^1\mathrm{A}'$ and (1)~$^3\mathrm{A}'$, as a function of $R$ for different configurations of CaF+CaF under the rigid rotor approximation. The effect of nonrigidity in SOC is also investigated and is presented by the green dashed and blue dashed-dotted curves. Lower panel:  energy difference between these two states for the corresponding geometries.  \label{fig:so}}
\end{figure}

\subsection{Spin-orbit and spin-spin interactions}
\label{subsec:SO}
Spin-orbit coupling (SOC) plays an important role in facilitating transitions between quantum states of different spin multiplicities in a molecular system, especially near the curve crossings. In the absence of SOC, electronic states with varying spin multiplicities remain uncoupled due to spin conservation, even when their potential energy surfaces intersect. Here we investigate the SOC between the ground state, (1)~$^1\mathrm{A}'$, and the lowest spin-polarized state, (1)~$^3\mathrm{A}'$, of the Ca$_2$F$_2$ dimer. This coupling may be useful for analyzing the collisional loss on the spin-polarized surface of CaF+CaF.

\begin{figure}[t]
\includegraphics[width=\linewidth]{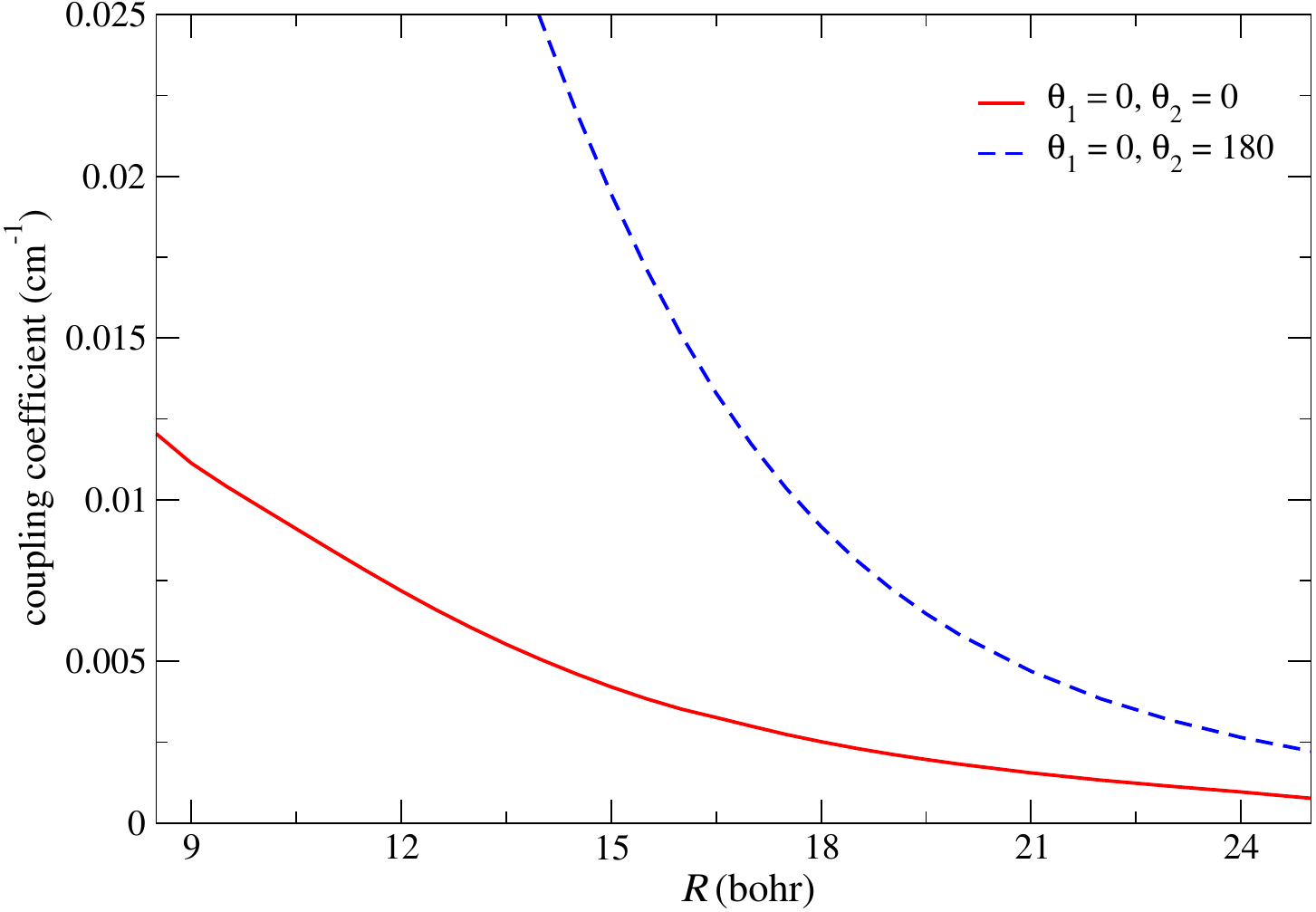}
\caption{The red solid curve represents the variation of the spin-spin coupling as a function of $R$ for a linear configuration, ($\theta_1=0^\circ, \theta_2=0^\circ$) under the rigid rotor approximation, whereas the blue dashed curve present the same for the second linear configuration, ($\theta_1=0^\circ, \theta_2=180^\circ$). For the latter, it is difficult to get a smooth curve at a very short range regime due to the issue of convergence. \label{fig:ss}}
\end{figure}

In Fig.~\ref{fig:so} we present one-dimensional cuts of the spin-orbit coupling matrix elements as a function of the center-of-mass distance. The SOC between the (1)~$^1\mathrm{A}'$ and (1)~$^3\mathrm{A}'$ states is zero asymptotically due to the lack of orbital angular momentum in the ground state of the CaF monomers. Conversely, at shorter intermonomer separations, the SOC becomes nonzero due to the admixture of electronically excited states with higher angular momentum to the ground state. This mixing is mediated by anisotropic intermolecular interactions. However, for linear geometries of CaF+CaF, the SOC remains zero. In such symmetric configurations, the (1)~$^1\mathrm{A}'$ and (1)~$^3\mathrm{A}'$ states correlate with $^1\Sigma$ and $^3\Sigma$, respectively, and the spin-orbit operator cannot couple two $\Sigma$ states due to symmetry constraints. In addition, the SOC for the global-minimum geometries is close to zero due to the ionic bonding in the Ca$_2^+$F$_2^-$ dimer and a very small admixture of excited states to the ground state.

We find that the SOC matrix elements are on the order of 5~cm$^{-1}$ or less near the minimum of the (1)~$^3\mathrm{A}'$ potential energy surface, under the rigid-rotor approximation for the CaF molecule, in a representative CaF+CaF configuration ($\theta_1 = 45^\circ$, $\theta_2 = 45^\circ$).

Notably, the magnitude of SOC remains largely unchanged even when the CaF bond length is elongated by more than 10\%, indicating that deviations from the rigid-rotor approximation do not significantly affect SOC in this system. We also observe finite SOC values for other geometries, including two alternative T-shaped arrangements and a parallel configuration of CaF+CaF. In all these cases, the SOC matrix elements remain below 5~cm$^{-1}$ near the minimum.

The lower panel of Fig.~\ref{fig:so} shows the energy difference between the (1)~$^1\mathrm{A}'$ and (1)~$^3\mathrm{A}'$ states  for all configurations where SOC is computed. In most geometries, this difference in energy exceeds 1000~cm$^{-1}$ that is two orders of magnitude larger than SOC. These results indicate that the effective spin-orbit-mediated coupling between the (1)~$^1\mathrm{A}'$ and (1)~$^3\mathrm{A}'$ states is very weak and may have a negligible effect on transitions between these two states.

Another coupling mechanism relevant for the spin-polarized surface, (1)~$^3\mathrm{A}'$, arises from the electron spin-spin interaction, and the corresponding effective Hamiltonian for the total electronic spin $\hat{S}$, $D(\hat{S}^2_z-\frac{1}{3}\hat{S}^2)$, can be parameterized by the zero-field splitting parameter $D$~\cite{KarmanPRA2023}. It has two contributions: magnetic dipole-dipole interactions and second-order spin-orbit coupling, and is approximated by $2(\alpha^2/R^3-\lambda(R))$. The direct magnetic dipole-dipole interactions can be approximated by $2\alpha^2/R^3$ \cite{StoofPRA1988}, where $\alpha$ is the fine structure constant, and $\lambda (R)$ is the second-order spin-orbit coupling \cite{TakekoshiPRA2012}. In Fig.~\ref{fig:ss} we present the variation of the spin-spin coupling $D$ for the two linear representative geometries of CaF+CaF. We note that the spin-spin coupling near the potential minimum of the (1)~$^3\mathrm{A}'$ state is of the order of $10^{-2}$~cm$^{-1}$, comparable to the spin-polarized $a~^3\Sigma^+$ state of the alkali-metal dimers \cite{KarmanPRA2023,TakekoshiPRA2012}. 

\subsection{Electric transition dipole moments}
\label{sub:TDM}
\begin{figure}[t!]
\includegraphics[width=\linewidth]{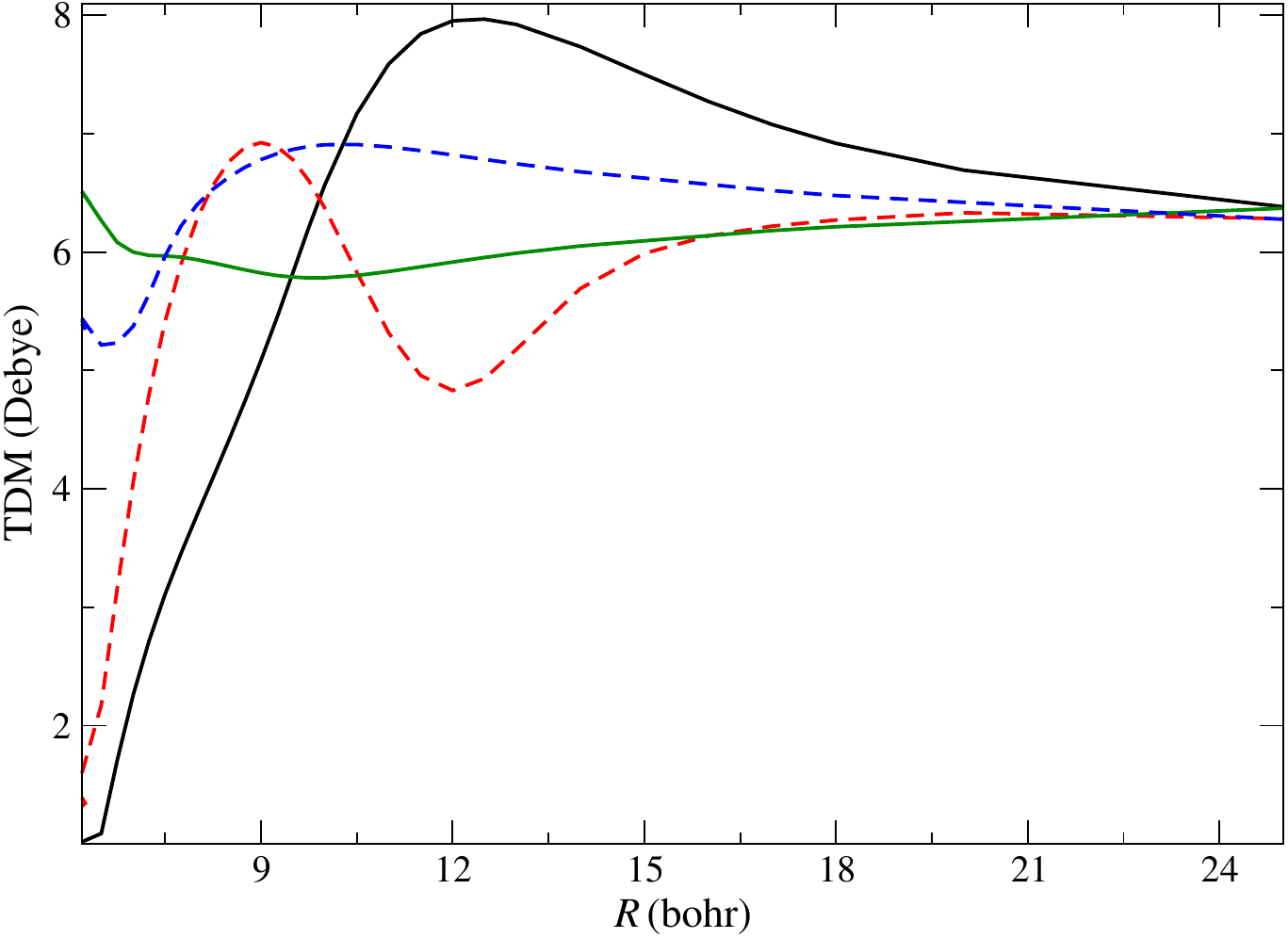}
\caption{Variation of electric transition dipole moments as a function of $R$ for the two alternative T-shaped configurations of CaF+CaF. The solid black and dashed red curves represent the variation of TDM for the (1)$^3\mathrm{A}'\rightarrow(2)^3\mathrm{A}'$ and (1)$^3\mathrm{A}'\rightarrow(3)^3\mathrm{A}'$ transitions, respectively, for the T-shape orientation ($\theta_1 = 0^\circ$, $\theta_2 = 90^\circ$). Similarly, the solid green and dashed blue curves show the same transitions for an alternative T-shaped configuration ($\theta_1 = 90^\circ$, $\theta_2 = 0^\circ$).  \label{fig:tdm}}
\end{figure}

\begin{figure*}
\includegraphics[width=\linewidth]{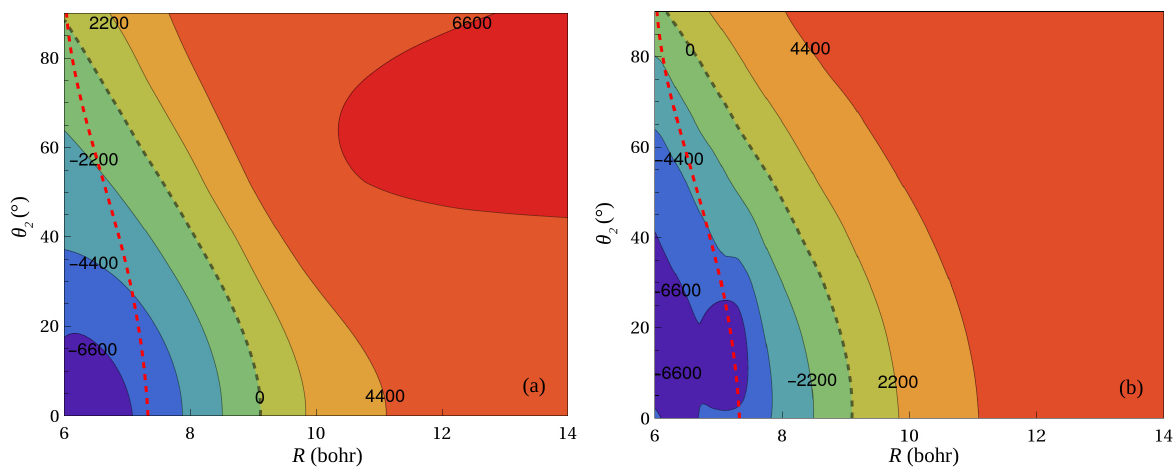}
\caption{Two-dimensional energy difference between the PES of the (1)~$^3\mathrm{A}'$ state shifted by 1064 nm of laser light and those of the (2)~$^3\mathrm{A}'$ and (1)~$^3\mathrm{A}''$ states are shown in panels (a) and (b), respectively. The energy difference is expressed in cm$^{-1}$. The crossing between the potential energy surfaces is indicated by the black dashed lines. The red dashed curves indicate the classical turning point for PES of the (1)~$^3\mathrm{A}'$ state. All the 2D PESs are computed under the rigid rotor approximation, with both $\theta_1$ and $\varphi$ set to zero. \label{fig:2d}}
\end{figure*}
The electric transition dipole moment (TDM) is a fundamental property of a molecular system, as it determines the rate of photoexcitation transitions. A large TDM between two quantum states enables stronger interaction with an external laser field, leading to a higher transition probability during the photoexcitation process. In this work, we present TDM results for the T-shaped geometries of the CaF+CaF complex to illustrate the variation of dipole-allowed transitions in a representative configuration.

In Fig.~\ref{fig:tdm} we present the absolute values of the electric transition dipole moment for the $(1)~^3\mathrm{A}'–(2)~^3\mathrm{A}'$ and $(1)~^3\mathrm{A}'–(3)~^3\mathrm{A}'$ transitions as a function of the center-of-mass distance. The calculations are performed using the MRCISD method for the two T-shaped configurations of CaF+CaF. Notably, the TDM between the (1)~$^3\mathrm{A}'$ and both (2)~$^3\mathrm{A}'$ and (3)~$^3\mathrm{A}'$ states is large across these orientations, consistent with the dipole allowed transitions between the X~$^2\Sigma^+$ and A~$^2\Pi$ electronic states of the CaF monomer, and the TDM converges to a constant value of 6.30 Debye. Additionally, the electric transition dipole moments from the (1)~$^3\mathrm{A}'$ state to another symmetry state, (1)~$^3\mathrm{A}''$, yield a large value, too (not presented here). The purpose of computing the TDM may be effective in analyzing the collisional loss on the spin-polarized triplet surface of CaF+CaF, facilitated by the laser-driven transition process, as discussed in the following subsection.

\subsection{Prospects for molecular collisional loss of CaF+CaF}
\label{subsec:loss}

The 1D-cut plots of the potential energy surfaces, spin-orbit coupling, and transition dipole moments are useful for understanding ultracold molecular collisional losses of CaF+CaF in their lowest spin-polarized state. We discuss it in the following. It is worth noting that the blue dashed curves in Fig.~\ref{fig:all} correspond to the PES of the spin-polarized (1)~$^3\mathrm{A}'$ state shifted by the presence of 1064 nm of light from the optical dipole trap. This shifted PES intersects several excited states, including (2)~$^3\mathrm{A}'$ and (1)~$^3\mathrm{A}''$, where these PESs exhibit attractive behavior. This confirms that these excited states can be reached by 1064 nm of light. 

Additionally, we compute the two-dimensional potential energy surfaces using the MRCISD method and make difference plots between the shifted PES of the (1)~$^3\mathrm{A}'$ state and (2)~$^3\mathrm{A}'$ and (1)~$^3\mathrm{A}''$ states. In Fig.~\ref{fig:2d} we present the corresponding 2D difference plots. In both cases, the energy difference between the two PESs passes through zero, indicating the existence of curve crossings between the PESs for several configurations of CaF+CaF. It further confirms that the excited states are accessible under a 1064 nm laser. Additionally, strong electric transition dipole moments exist for these excited states due to the dipole allowed CaF ($^2\Sigma \rightarrow ^2\Pi$) transition. These calculations do not directly confirm the formation of a four-body collision complex on the spin-polarized surface of CaF+CaF. But, the presence of near-resonant PES crossings and strong transition dipole moments suggests that, if such a complex were to form during a collision, it could undergo laser-induced transitions to electronically excited states. This pathway may serve as a contributing factor to collision loss mechanisms on the spin-polarized surface of CaF+CaF, as is the case in
alkali dimers \cite{BausePRL2023}.

Alternatively, the collisional loss for the spin-polarized surface of CaF+CaF may occur by a different mechanism: a transition from the triplet surface to the reactive singlet surface of CaF+CaF facilitated by spin-orbit coupling results in the chemical reaction. This transition will be effective when the potential energy surfaces of the two states closely approach or directly cross each other, particularly in the presence of strong spin-orbit coupling between the two states. However, we compute several one-dimensional cuts for the PESs of CaF+CaF, but we can not observe any direct curve crossing between the (1)~$^1\mathrm{A}'$ and (1)~$^3\mathrm{A}'$ states. Additionally, we find that the SOC between these states is of the order of 5~cm$^{-1}$, while the energy difference is of the order of 1000~cm$^{-1}$. This relatively large energy gap may suppress the transition process from the spin-polarized triplet surface to the reactive singlet surface. While these findings suggest that such a transition is unlikely under the conditions considered here, the potential contribution of spin-orbit–mediated transitions to collisional loss cannot be completely ruled out without a scattering calculation and taking into account sticking dynamics and lifetime~\cite{SardarJPCA2023}.

In addition, the spin-spin coupling for the lowest spin-polarized (1)~$^3\mathrm{A}'$ state is comparable to the $^3\Sigma$ state of light alkali-metal dimers, since in both cases it is dominated by the magnetic dipole-dipole interaction between valence electrons. This coupling strength was sufficient to account for the observed background loss rate in Na+NaLi collisions, which is more than an order of magnitude smaller than the universal loss rate constant \cite{ParkPRX2023}. By analogy, spin-spin coupling in the (1)~$^3\mathrm{A}'$ state may play a significant role in the loss processes of CaF+CaF collisions on the spin-polarized surface at ultracold temperatures. However, all these remarks about the collisional loss, of course, remain speculative at this point and will have to be validated by detailed dynamical calculations, which are beyond the scope of the present study.

\section{Conclusions}
\label{sec:summary}
We investigated the intermolecular interactions between two laser-cooled CaF molecules in both their electronically ground and excited states. Using state-of-the-art \textit{ab initio} quantum chemistry methods, we computed one-dimensional cuts of potential energy surfaces under the rigid rotor approximation, considering excitation energies up to approximately 19000~cm$^{-1}$ for the CaF monomer. Our analysis revealed that molecular nonrigidity has a negligible impact on the shape and topology of the PESs for the Ca$_2$F$_2$ dimer. We also calculated the spin-orbit coupling matrix element between the two lowest electronic states, (1)~$^1\mathrm{A}'$ and (1)~$^3\mathrm{A}'$, and the spin-spin coupling within the (1)~$^3\mathrm{A}'$ state. In addition, we determined the electric transition dipole moments between the states: (1)~$^3\mathrm{A}'$ to (2)~$^3\mathrm{A}'$, and (1)~$^3\mathrm{A}'$ to (3)~$^3\mathrm{A}'$. 

Finally, by analyzing the excited-state PESs, spin-dependent couplings, and transition dipole moments, we proposed two possible mechanisms for the observed collisional losses on the spin-polarized surface of CaF+CaF \cite{CheukPRL2020}. One mechanism involves spin-orbit-induced transitions to the reactive singlet surfaces, while the other pertains to photoexcitation-induced losses. In the latter case, collisions between two CaF molecules on the lowest spin-polarized surface may result in the formation of a transient four-body collision complex, which can undergo laser-induced electronic transitions to excited states, ultimately leading to trap loss. Notably, our \textit{ab initio} calculations of excited-states PESs suggest that the photoexcitation-induced process may be one of the important sources of collisional losses observed for the spin-polarized surface of CaF+CaF, as was the case for alkali-metal molecules discussed in Ref.~\cite{ChristianenPRL2019}. Although spin-orbit-mediated transition from the spin-polarized surface to the ground reactive surface appears unlikely under the conditions considered in this study, their contribution to collisional loss can not be entirely ruled out. Additionally, the spin-spin coupling on the spin-polarized state, (1)~$^3\mathrm{A}'$, is significant and comparable to that in the alkali dimer \cite{ParkPRX2023}, and may also play a role in the observed losses in CaF+CaF collisions on the spin-polarized surface. Although these remarks remain speculative and can be validated by full-dimensional quantum scattering calculations.

\begin{acknowledgements}
D.S.~is thankful for initial discussions with R. Maitra at IITB. We gratefully acknowledge the National Science Centre Poland (Grant No.~2020/38/E/ST2/00564) and the European Union (ERC, 101042989 -- QuantMol) for financial support and the Poland’s high-performance computing infrastructure PLGrid (HPC Centers: ACK Cyfronet AGH) for providing computer facilities and support (computational Grant No.~PLG/2024/017844). JLB acknowledges funding from AFOSR-Multidisciplinary University Research Initiatives Grant No. GG016303.
\end{acknowledgements}

\section*{DATA AVAILABILITY}
The data supporting this study’s findings are available upon request.
request.

\bibliography{sample}

\end{document}